\documentclass[
reprint,
amssymb,
aps,
prl,
floatfix,
superscriptaddress]{revtex4-1}

\usepackage{graphicx}
\usepackage{dcolumn}
\usepackage{bm}
\usepackage{epstopdf}
\usepackage{braket}
\usepackage{soul}
\usepackage{amsmath}
\usepackage[euler]{textgreek}

\begin{document}

\title{Ramsey-Bord\'e matter-wave interferometry for laser frequency stabilization at $10^{-16}$ frequency instability and below}

\author{Judith Olson}
	\email{judith.olson@colorado.edu}
	\affiliation{Time and Frequency Division, National Institute of Standards and Technology, Boulder, Colorado 80305, USA}
	\affiliation{Department of Physics, University of Colorado, Boulder, Colorado 80309, USA}

\author{Richard W. Fox}
	\affiliation{Time and Frequency Division, National Institute of Standards and Technology, Boulder, Colorado 80305, USA}

\author{Tara M. Fortier}
	\affiliation{Time and Frequency Division, National Institute of Standards and Technology, Boulder, Colorado 80305, USA}

\author{Todd F. Sheerin}
	\affiliation{Department of Aeronautics and Astronautics, Massachusetts Institute of Technology, Cambridge, Massachusetts 02139}
	\affiliation{The Charles Stark Draper Laboratory, Inc., Cambridge, Massachusetts 02139, USA}

\author{Roger C. Brown}
	\affiliation{Time and Frequency Division, National Institute of Standards and Technology, Boulder, Colorado 80305, USA}
	
\author{Holly Leopardi}
	\affiliation{Time and Frequency Division, National Institute of Standards and Technology, Boulder, Colorado 80305, USA}
	\affiliation{Department of Physics, University of Colorado, Boulder, Colorado 80309, USA}

\author{Richard E. Stoner}
	\affiliation{The Charles Stark Draper Laboratory, Inc., Cambridge, Massachusetts 02139, USA}

\author{Chris W. Oates}
	\affiliation{Time and Frequency Division, National Institute of Standards and Technology, Boulder, Colorado 80305, USA}

\author{Andrew D. Ludlow}
\email{andrew.ludlow@nist.gov}
	\affiliation{Time and Frequency Division, National Institute of Standards and Technology, Boulder, Colorado 80305, USA}
	\affiliation{Department of Physics, University of Colorado, Boulder, Colorado 80309, USA}
	
\date{\today}

\begin{abstract}
	
We demonstrate Ramsey-Bord\'e (RB) atom interferometry for high performance laser stabilization with fractional frequency instability $<2 \times 10^{-16}$ for timescales between 10 and 1000s. The RB spectroscopy laser interrogates two counterpropagating $^{40}$Ca beams on the $^1$S$_0$ -- $^3$P$_1$ transition at 657 nm, yielding 1.6 kHz linewidth interference fringes. Fluorescence detection of the excited state population is performed on the (4s4p) $^3$P$_1$ -- (4p$^2$) $^3$P$_0$ transition at 431 nm. Minimal thermal shielding and no vibration isolation are used. These stability results surpass performance from other thermal atomic or molecular systems by one to two orders of magnitude, and further improvements look feasible.

\end{abstract}

\maketitle
\setuldepth{.}

Lasers with exceptional frequency stability play a foundational role in fundamental physics, precision measurement, and technological applications. Noteworthy is the optical clock, which derives its unmatched measurement precision from a frequency-stabilized laser locked to a narrowband atomic resonance \cite{Nemitz2016,huntemann2016single}. Such precision enables searches for both dark matter and variations of fundamental constants \cite{stadnik2016enhanced,derevianko2014hunting,safronova2018search} while motivating redefinition of the International System (SI) second \cite{riehle2015towards}. Ultra-stable lasers are key components in many tests of relativity \cite{ashby2018null, georgescu2017special,chou2010optical} and  lie at the heart of gravitational wave observatories \cite{abbott2016observation,acernese2014advanced,Canuel2018}. They are used in quantum control and manipulation, including recent explorations of SU($n$) interactions and quantum simulation \cite{zhang2014spectroscopic,schmidt2003realization}. Furthermore, their stability can be readily transferred \cite{leopardi2017single,Xie2016} across the electromagnetic spectrum for advanced communication and radar systems or next-generation position, navigation, and timing.

We present a method for ultra-precise laser-frequency stabilization based on Ramsey-Bord\'e (RB) matter-wave interferometry using a thermal calcium beam. This approach enables frequency instability $< 2 \times 10^{-16}$ and possesses a vibration sensitivity that could be orders of magnitude smaller than the standard workhorse of optical frequency stabilization, the Fabry-Perot cavity. Such cavities typically employ high quality factor ($Q$) and signal-to-noise ($S/N$), offering impressive 10$^{-16}$ instability at short times, with recent progress on very long \cite{hafner20158} and cryogenic \cite{matei20171,robinson2019crystalline} cavities reaching into the 10$^{-17}$ decade. This performance level is important for minimizing the Dick effect in optical clocks and enabling coherent interrogation of ultra-long-lifetime atomic states \cite{jiang2011making}. Yet increasing challenges of further reducing cavity thermal noise \cite{numata2004thermal} have spurred research in alternative approaches aimed at circumventing the thermal noise problem, including superradiant lasers \cite{norcia2018frequency,schaffer2017towards,wang2012superradiant} and spectral hole burning \cite{cook2015laser}.

Here, we employ optical Ramsey techniques \cite{bergquist1977saturated, barger1979resolution} common to atom interferometry and precision measurements \cite{borde1989atomic, kovachy2015quantum,hamilton2015atom,dubetsky2016atom}. Our specific design builds upon the experimental geometry laid out in \cite{borde1984optical}. This technique was chosen for its ability to resolve narrow spectral features from a high-flux thermal beam of atoms or molecules despite their large velocity distribution. In this way, the traditionally problematic trade-offs in atomic systems between $Q$ and $S/N$ can be avoided, enabling both relatively fast and precise frequency corrections. To our knowledge, this is the first time a thermal frequency reference has demonstrated stability at $\leq 10^{-15}$, highlighting an entirely new regime of performance. This unlocks new applications for thermal ensembles, which are simpler than cold-atom systems and generally possess superior long-term stability and accuracy over mechanical references such as cavities.

\begin{figure}[ht]
	\includegraphics[width=0.475\textwidth]{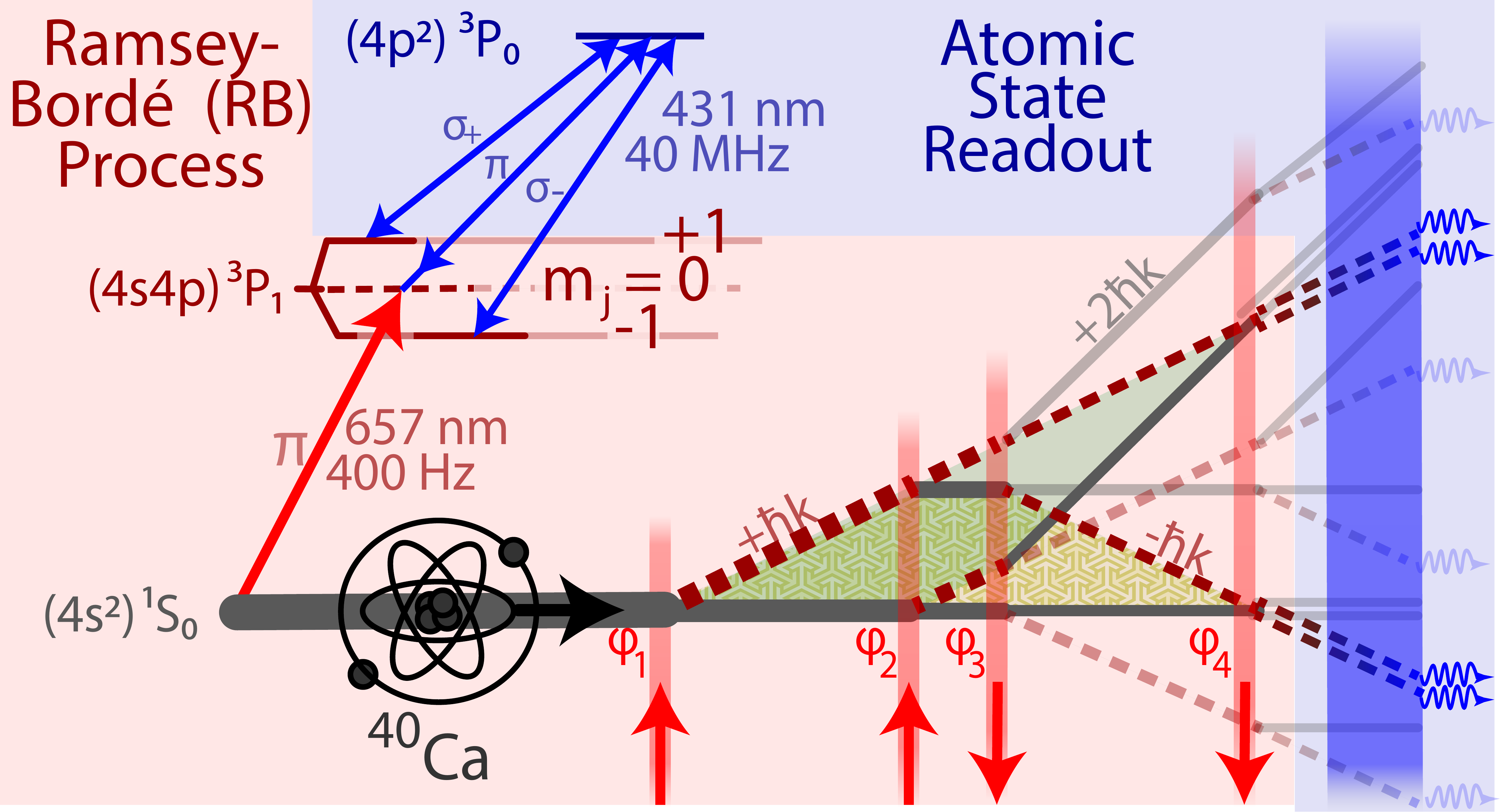}
	\centering
	\caption{\label{levelsandinterferometer} Relevant electronic levels and transitions of neutral $^{40}$Ca. Also shown is the quantum trajectory of a calcium atom passing through the RB interferometer.  The atom undergoes four laser interactions with phase $\phi_i$, which repeatedly split the atomic momentum state. Two possible closed trajectories exist (filled areas), creating matter-wave interference. The $^3$P$_1$ -- $^3$P$_0$ transition is then rapidly cycled, and the scattered 431 nm photons collected.
	}
	
\end{figure}

Ramsey-Bord\'e interferometry \cite{borde1984optical} employs four $\pi / 2$ laser-atom interactions that repeatedly split the atomic wave function, resulting in two closed quantum trajectories (Fig.~\ref{levelsandinterferometer}) that encode interferometric fringes on the atomic state populations \cite{borde1989atomic}. The excited state probability, $P_e$, is a periodic function of the interrogation laser frequency, $\nu$, around the natural transition frequency, $\nu_0$, given by the energy difference between atomic states, as well as the phase difference, $\Phi = \phi_2 - \phi_1 + \phi_4-\phi_3$, accumulated from each sequential laser interaction of phase $\phi_i$ (Fig. \ref{levelsandinterferometer}) such that:

\begin{align}
\label{RBeqn}
P_e  \propto  \cos \Bigl[ 4\pi  (\nu - \nu_{0}\pm\delta\nu_{recoil}  + \nu_{sys}) T + \Phi \Bigr],
\end{align}

\begin{figure}[ht]
 	\includegraphics[width=0.475\textwidth]{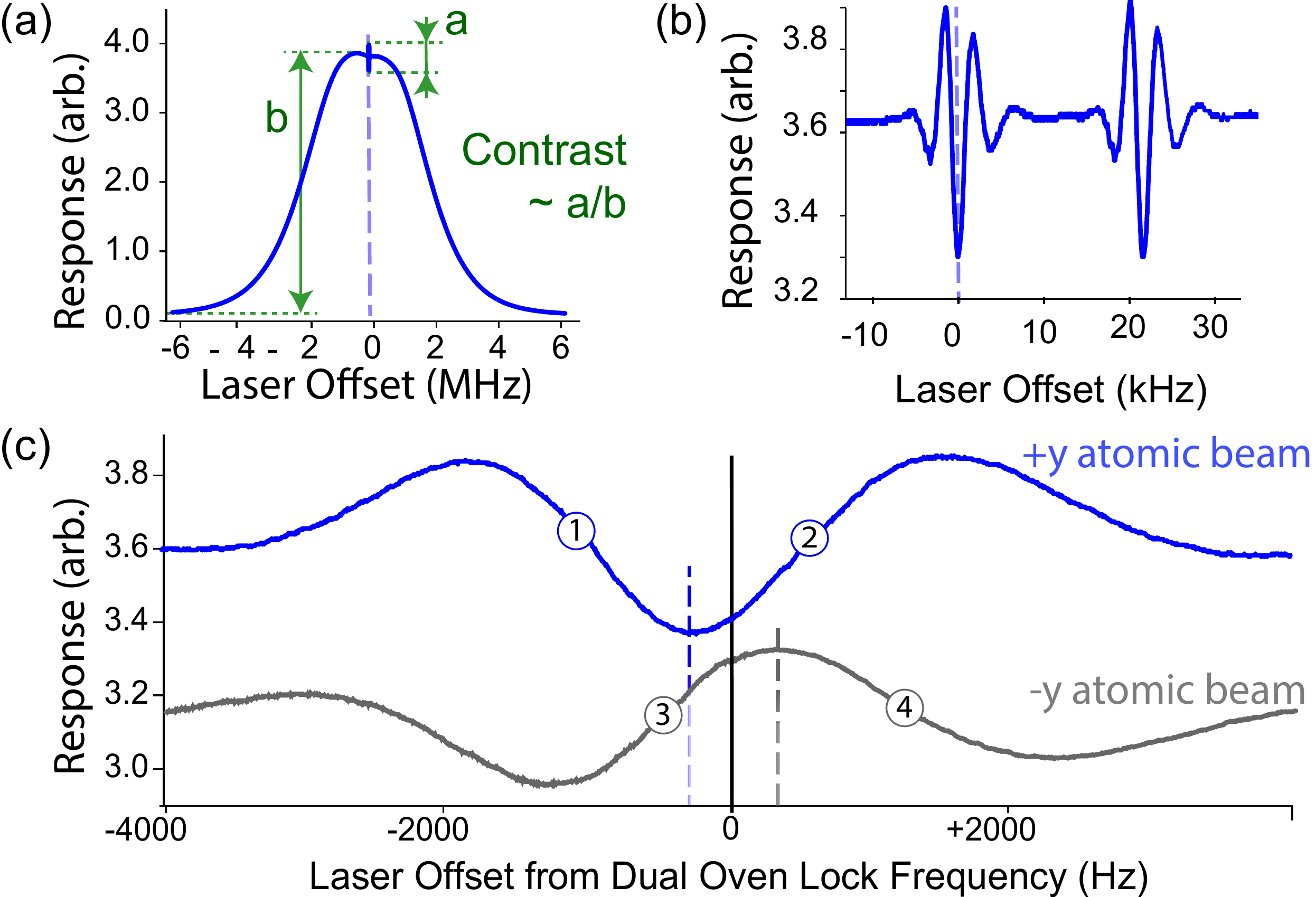}
 	\centering
 	\caption{\label{fringes} Typical RB fringes for a range of frequency sweeps. (a) MHz scale Doppler profile of one atomic beam response, with accompanying measure of fringe contrast (b) Mid-range sweep displaying two RB fringe components from a single atomic beam (c) Simultaneous fine sweep across the RB fringes from both counterpropagating atomic beams. Dashed lines indicate extrema feature frequencies, and points 1 to 4 indicate four locking frequency locations.
 	}
 	
\end{figure}

\noindent where $T$ is the free-evolution time (i.e. Ramsey time) between the first and second, or equivalently, the third and fourth laser interactions. The atomic recoil frequency is $\delta\nu_{recoil} = h/{2 m \lambda^2} \cong 11.5 \ \text{kHz}$ for the $^1$S$_0$ -- $^3$P$_1$ transition in calcium, where $h$ is Planck's constant, $\lambda$ is the transition wavelength, and $m$ is the atomic mass. All other systematic frequency shifts, such as those caused by stray fields or time dilation, are included in $\nu_{sys}$. While the RB technique is ideally free from first-order Doppler shifts, residual first-order Doppler effects are traditionally included in the fringe phase, $\Phi$. $T$ varies across the ensemble's broad Maxwell-Boltzmann velocity distribution \cite{barger1981influence}, yielding only a few resolvable RB fringe periods, as seen in Fig.~\ref{fringes}.

\begin{figure}[ht]
	\includegraphics[width=0.485\textwidth]{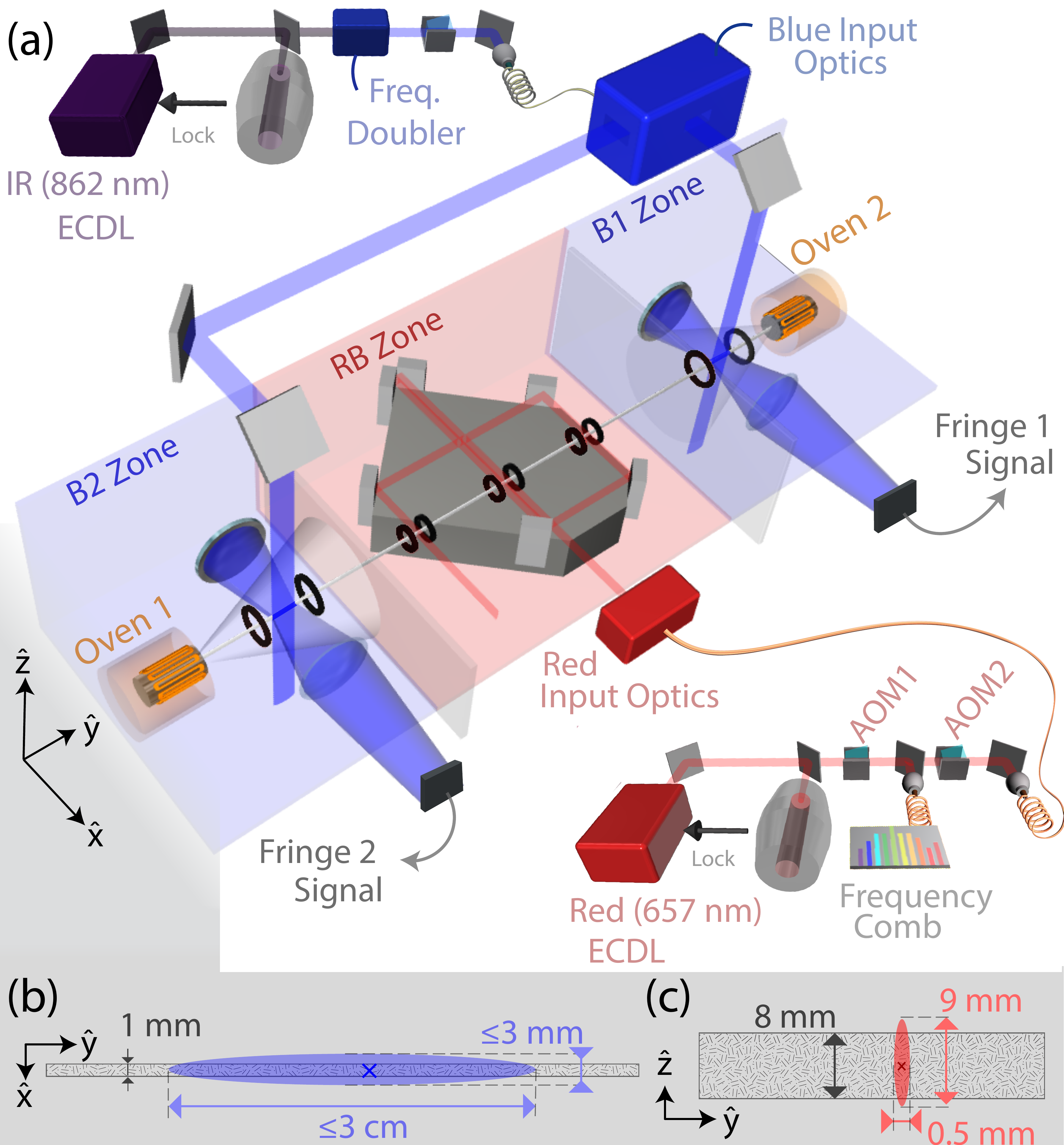}
	\centering
	\caption{\label{setup} Experimental setup. (a) Both 862 nm (later frequency doubled to 431 nm) and 657 nm lasers are cavity pre-stabilized. Input optics mounted on the vacuum chamber shape and deliver the light. The vacuum chamber is partitioned into two blue detection zones (B1 and B2) and the RB zone. Within vacuum, the interrogation laser is steered by a monolithic glass assembly to intersect the atomic beam at four locations. The distance between beams 1 and 2 (or 3 and 4) is 9 cm, whereas the spacing between beams 2 and 3 is roughly 1 mm.  In regions B1 and B2, exited-state atoms are rapidly cycled while their scatter is focused onto an external detector. Magnetic fields are indicated by black coils. Apertures between zones collimate the atomic beams.  AOM: acousto-optic modulator, ECDL: external cavity diode laser. (b,c) Approximate blue (431 nm) and red (657 nm) laser profiles, respectively, are given relative to the atomic beam dimensions.  Typical laser powers are 10 mW (431 nm) and 25 mW (657 nm), incident on the atoms.}
	
\end{figure}

Two thermal beams of calcium atoms are used for frequency correction. These counterpropagating beams are defined by the same apertures to precisely overlap (Fig. \ref{setup}). Atoms from opposing ovens experience the interferometer in reverse orders, causing the total accumulated phase, $\Phi$, of each oven's fringe to be nominally equal and opposite. Therefore, locking the laser frequency to the bi-directional average reduces sensitivity to residual first-order Doppler shifts from optical path length variation, imperfect interferometer alignment, and RB laser input angle. Though a technique reversing laser propagation on a single atomic beam was also employed \cite{Ito1994}, we observed fringes with more robust anti-symmetry from dual atomic beams, ultimately yielding better laser stability.

The fundamental vibrational insensitivity of the RB interferometer can be appreciated through comparison to a Fabry-Perot cavity.  Thermo-mechanical noise and acceleration-induced cavity deformation often limit the stability of the Fabry-Perot cavity, causing an optical path length change, ${\delta l}$, corresponding to a fractional frequency shift $\frac{\delta \nu}{\nu} = \frac{\delta l}{l_{cav}}$, for cavity length $l_{cav}$.  For the RB interferometer, a similar change in optical path yields a phase difference, $\Delta \Phi = 2\pi \frac{\delta l}{\lambda}$.  From Eq. (\ref{RBeqn}), the resulting fractional frequency shift is $\frac{\delta \nu}{\nu} = (\frac{\delta l}{\lambda}\frac{1}{2T})/\nu = \frac{\delta l}{l_{RB}}\frac{v_{atom}}{2c}$, where $v_{atom}$ ($l_{RB}$) is the atomic velocity (distance traveled) during $T$.  For the reasonable approximation $l_{cav} \approx l_{RB}$ (here $l_{RB}=9$ cm), optical-path-length driven frequency shifts to the RB system are reduced by a factor $v_{atom}/2c \sim 10^{-6}$ compared to the Fabry-Perot cavity.  On the other hand, because atoms in a RB interferometer are free bodies, accelerations alter their trajectory relative to the laser field, introducing a frequency shift sensitivity predominantly along the laser propagation axis \cite{Wilpers2003}, $\frac{\delta \nu}{\nu}=\frac{l_{RB} a}{v_{atom} c}$ for acceleration $a$.  For parameters listed above, this corresponds to a fractional-frequency acceleration sensitivity of $\frac{\delta \nu}{\nu}=5 \times 10^{-12} a/g$, for gravitational acceleration $g$.  While this acceleration sensitivity is already competitive with the best Fabry-Perot cavity designs \cite{Webster2011,Leibrandt2013}, the atomic beam reversal technique employed here allows further sensitivity reduction.


A natural choice of quantum absorber for RB-interferometer-based laser stabilization is calcium, using the $^1$S$_0$ -- $^3$P$_1$ clock transition at 657 nm (Fig. \ref{levelsandinterferometer}) with an excited state lifetime near 400 \textmu s \cite{Wilpers2003,McFerran2010,Shang2017}. Because the RB fringe linewidth decreases as $1/T$, an ideal RB interferometer interrogates the atoms for a time approaching the natural lifetime. For calcium effusing from an oven near 625 $^\circ$C with average speed $>$600 m/s, this requires an interferometer length of $\sim20$ cm, well-suited to a compact apparatus.

Additionally, $^{40}$Ca possesses a relatively strong transition from the RB excited state to a doubly-excited state, (4s4p) $^3$P$_1$ -- (4p$^2$) $^3$P$_0$ at 431 nm (Fig.~\ref{levelsandinterferometer}). A second laser resonantly applied to this transition is used for electron-shelving fluorescence detection \cite{nagourney1986shelved} of the RB fringes, enabling hundreds of photons to be emitted per atom without appreciable optical pumping into dark states. Both the 657 nm and 431 nm sources are accessible with commercially available diode lasers, with sufficient power to efficiently drive these transitions.

The calcium apparatus uses a vacuum chamber with pressure near 10$^{-5}$ Pa, which is partitioned symmetrically into three zones (Fig. \ref{setup}). The intensity- and phase-stabilized 657 nm interrogation laser is delivered to the RB zone through a viewport and then directed along a nearly meter-long folded path in vacuum with exacting parallelism. This is accomplished via an ultra-low-expansion (ULE) glass spacer with angular tolerances $\sim$1 arcsecond. The spacer sits below beam-line with optically contacted high-reflection ULE mirrors extending beyond the plane of the interrogation laser. Also in the central RB zone, the magnetically insensitive $^3$P$_1$ (m$_\text{J}$ = 0) sublevel is spectrally isolated from m$_\text{J}$ = $\pm$1 with  $\approx$700 \textmu T fields during RB interrogation.

Elsewhere, laser light at 431 nm is intensity-stabilized, shaped, and split into two paths for delivery to the blue detection zones (B1 and B2 in Fig. \ref{setup}).  Each of these zones employs a single Helmholtz coil pair to generate an $\sim$800 \textmu T field over the approximately 3 cm long laser-atom interaction (Fig. \ref{setup}b).  Of the hundreds of photons scattered by each excited-state atom, approximately 10\% are imaged onto a detector (Fig. \ref{setup}).  The blue laser polarization incident on the atomic beam is adjusted to optimize the collected laser-induced fluorescence.

Both laser beam profiles at 657 nm and 431 nm, as well the atomic beam profile, are steeply elongated in one dimension. The atomic beam is shaped by fine apertures before and after the RB zone, restricting its profile to 1 $\times$ 8 mm. The laser profiles are shaped using anamorphic prism pairs. All relevant beam dimensions are given in Fig. \ref{setup}b,c. The red beam's narrow waist increases transit time broadening at each RB interaction zone to address a larger range of velocity classes without appreciable impact on the fringe linewidth. However, the short Rayleigh range prevents uniform collimation throughout the in-vacuum optical path.  Taken together with wavefront imperfections and inhomogeneous pulse area from the thermal beam velocity distribution, fringe contrast is reduced.  In theory, contrast can reach 25\%, though optimized values between 13-18\% were observed.  Interestingly, contrast is typically maximized when the interrogation laser is either slightly convergent or divergent.

Frequency corrections are generated by monitoring the fluorescence amplitude at four distinct interrogation laser frequencies - two for each fringe feature (Fig. \ref{fringes}c). The frequencies are modulated with an acousto-optic-modulator (AOM2 in Fig. \ref{setup}). For each oven the corresponding fluorescence levels are sampled and compared, and the two-oven average generates the frequency corrections via AOM1 (Fig.~\ref{setup}). The frequency difference between the two fringe extrema is tracked to accommodate relative phase variations between atomic beams. A single experimental cycle, including fluorescence collection, AOM updates, and 4 ms of dead time, is 10-30 ms.

\begin{figure}[ht]
	\includegraphics[width=.475\textwidth]{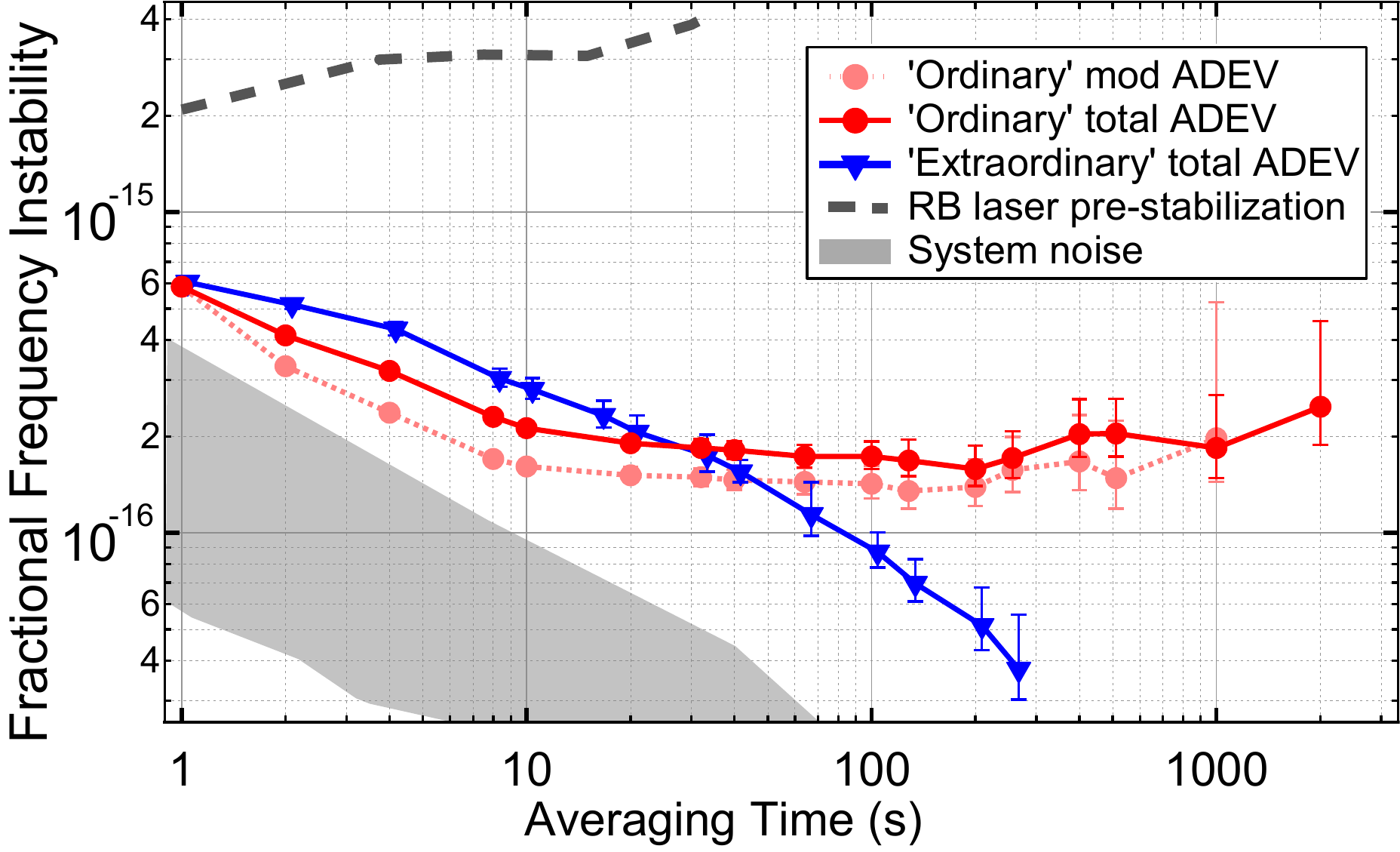}
	\centering
	\caption{\label{ADevPlot} Allan deviations for two calcium-stabilized frequency measurements with 1$\sigma$ error bars. The general system performance is well represented by the data set labeled `Ordinary' (red circles), whereas `Extraordinary' (blue triangles) highlights shorter data segments exhibiting exceptional performance, averaging below $10^{-16}$. The RB laser pre-stabilization is shown after a 2 Hz/s drift removal, though this drift is present during calcium system operation. The measured range of frequency instability contributions from detection noise sources falls within the gray region, see text.}
	
\end{figure}

The RB-stabilized laser instability is assessed via a frequency comb referenced to a ytterbium optical lattice clock system with instability $<10^{-16}$ after averaging $\approx$10 s \cite{mcgrew2018atomic}. On shorter times, the Yb clock instability is largely derived from an advanced FP cavity with instability $\leq 2\times10^{-16}$ at 1 s. Utilizing these two stable references and the low noise frequency comb ensures meaningful measurements of the $^{40}$Ca stabilized laser frequency for all time scales presented here.

We evaluate the frequency stability of the system for data sets taken over two time scales: one `ordinary' set over 1.4 hours and one `extraordinary' set over 10 minutes. Both are plotted in Fig. \ref{ADevPlot}, where the red curves represent typical (`ordinary') performance, which employs basic temperature control of the apparatus at the $\sim10$ mK level. Instability near $2\times 10^{-16}$ could be maintained for time scales of an hour, during which a flicker noise floor dominates around 1.2$\times$10$^{-16}$. This data includes a $\sim$100 \textmu Hz/s measured drift. As laser stability is sometimes reported with the modified ADEV stability estimator, this is also displayed in Fig. \ref{ADevPlot}.

The `extraordinary' blue curve of Fig.~\ref{ADevPlot} averages into the 10$^{-17}$ decade over $\sim$10 minutes of frequency measurements. Similar instability can be seen for periods up to 20 minutes, usually after the apparatus has run continuously for some time. This data set employed no temperature stabilization, and a linear frequency drift near 1 mHz/s was removed for the analysis. Averaging over longer intervals yields instability closer to that of the `ordinary' data in Fig. \ref{ADevPlot}.  In general, RB-stabilization with this system yields 1 s performance between 3-8$\times 10^{-16}$ with linear frequency drifts of $\leq1$ mHz/s.  Importantly, all stability levels presented were observed both with and without vibration isolation.

The gray region in Fig.~\ref{ADevPlot} denotes a range of detection noise, measured by tuning the 657 nm laser away from the RB fringes but still on the Doppler pedestal (Fig. \ref{fringes}a) and then scaling the fluctuations in detected fluorescence by the fringe discriminator slope, which yields the corresponding frequency fluctuations. This noise is a combination of photon shot noise and technical detector noise, and our measured frequency instability falls near this noise level for periods $<$10 s. At longer time intervals, we observed that temperature fluctuations of the apparatus typically correlate strongly with frequency wander.  We also investigated other sources of instability which were found to be less significant, including fluctuations in magnetic fields, laser intensity and polarization, laser pointing, vacuum levels, vibrations, atomic beam flux, and atomic collisions \cite{Olson2019}.

The exceptional instability demonstrated here compared to other thermal systems is facilitated by four important design considerations. First, the high atomic flux ($10^{10}$ to $10^{12}$ atoms per second) yields an atom shot-noise limited frequency instability $\leq 10^{-17}$ at short times, rivaling future thermal noise limits in cryogenically-cooled optical cavities. Second, dual ovens allow cancellation of fringe-phase ($\Phi$) drifts that otherwise degrade mid- and long-term performance. Third, enclosing the RB laser path within vacuum suppresses deleterious phase and pointing noise. Lastly, the excited-state cycling detection scheme used here is quite efficient - even after accounting for solid angle collection and detector quantum efficiency, most atoms contribute $\gg1$ photon to the resulting frequency correction.

A number of efforts look promising towards improving either the short-term stability (via improved detection system, atomic beam parameters, fringe contrast) or the long-term performance (via temperature stabilization).  For example, segmented RB interrogation using distinct pairs of phase-stabilized lasers could enhance fringe contrast closer to the fundamental limit of $25\%$, whereas simultaneous interrogation of both RB recoil components could yield a straightforward $\sqrt{2}$ enhancement in S/N.  Measurements of the RB system temperature sensitivity suggest that 1 mK stability could support stability at the $10^{-17}$ level.  This could make the calcium RB system compelling as an optical flywheel in a next-generation optical time scale, akin to the role played by hydrogen masers in conventional time scales.


However, the performance demonstrated here is already intriguing in the search for physics beyond the Standard Model.  Benefiting from relatively high-bandwidth and S/N, the calcium RB system enables searches for dark-matter interactions at mass and time scales beyond the reach of most optical clocks \cite{wcislo2017experimental,PhysRevLett.114.161301}. Furthermore, the simplicity and precision of our system could be exploited for isotope shift measurements in the search for new interactions via King's plot nonlinearities \cite{Berengut2018}.  Though the full potential of the RB system has not yet been seen, the instability results demonstrated here, combined with the system's low vibration sensitivity, atom-number scalability, and compact size, make RB optical-frequency stabilization a promising technique for a wide range of applications.

\begin{acknowledgments}
The authors would like to thank J.~Sherman, D.T.C.~Allcock, E.~de Carlos-Lopez, W.~Douglas, the NIST Yb lattice clock team (especially W.~McGrew and X.~Zhang), and all those whose research led us here. J.~O.~ and H.~L.~ would also like to acknowledge the NIST PREP cooperative agreement with the University of Colorado Boulder. T.~F.~S.~ acknowledges support from the Department of Defense (DoD) through the National Defense Science \& Engineering Graduate Fellowship (NDSEG) Program and from Draper through the Draper Fellow Program. We thank NIST, DARPA QuASAR, and DARPA STOIC for funding this research.

\end{acknowledgments}

\bibliographystyle{apsrev4-1}

\begin{thebibliography}{40}%
	\makeatletter
	\providecommand \@ifxundefined [1]{%
		\@ifx{#1\undefined}
	}%
	\providecommand \@ifnum [1]{%
		\ifnum #1\expandafter \@firstoftwo
		\else \expandafter \@secondoftwo
		\fi
	}%
	\providecommand \@ifx [1]{%
		\ifx #1\expandafter \@firstoftwo
		\else \expandafter \@secondoftwo
		\fi
	}%
	\providecommand \natexlab [1]{#1}%
	\providecommand \enquote  [1]{``#1''}%
	\providecommand \bibnamefont  [1]{#1}%
	\providecommand \bibfnamefont [1]{#1}%
	\providecommand \citenamefont [1]{#1}%
	\providecommand \href@noop [0]{\@secondoftwo}%
	\providecommand \href [0]{\begingroup \@sanitize@url \@href}%
	\providecommand \@href[1]{\@@startlink{#1}\@@href}%
	\providecommand \@@href[1]{\endgroup#1\@@endlink}%
	\providecommand \@sanitize@url [0]{\catcode `\\12\catcode `\$12\catcode
		`\&12\catcode `\#12\catcode `\^12\catcode `\_12\catcode `\%12\relax}%
	\providecommand \@@startlink[1]{}%
	\providecommand \@@endlink[0]{}%
	\providecommand \url  [0]{\begingroup\@sanitize@url \@url }%
	\providecommand \@url [1]{\endgroup\@href {#1}{\urlprefix }}%
	\providecommand \urlprefix  [0]{URL }%
	\providecommand \Eprint [0]{\href }%
	\providecommand \doibase [0]{http://dx.doi.org/}%
	\providecommand \selectlanguage [0]{\@gobble}%
	\providecommand \bibinfo  [0]{\@secondoftwo}%
	\providecommand \bibfield  [0]{\@secondoftwo}%
	\providecommand \translation [1]{[#1]}%
	\providecommand \BibitemOpen [0]{}%
	\providecommand \bibitemStop [0]{}%
	\providecommand \bibitemNoStop [0]{.\EOS\space}%
	\providecommand \EOS [0]{\spacefactor3000\relax}%
	\providecommand \BibitemShut  [1]{\csname bibitem#1\endcsname}%
	\let\auto@bib@innerbib\@empty

    \bibitem[{\citenamefont{Nemitz et~al.}(2016)\citenamefont{Nemitz, Ohkubo,
  Takamoto, Ushijima, Das, Ohmae, and Katori}}]{Nemitz2016}
\bibinfo{author}{\bibfnamefont{N.}~\bibnamefont{Nemitz}},
  \bibinfo{author}{\bibfnamefont{T.}~\bibnamefont{Ohkubo}},
  \bibinfo{author}{\bibfnamefont{M.}~\bibnamefont{Takamoto}},
  \bibinfo{author}{\bibfnamefont{I.}~\bibnamefont{Ushijima}},
  \bibinfo{author}{\bibfnamefont{M.}~\bibnamefont{Das}},
  \bibinfo{author}{\bibfnamefont{N.}~\bibnamefont{Ohmae}}, \bibnamefont{and}
  \bibinfo{author}{\bibfnamefont{H.}~\bibnamefont{Katori}},
  \bibinfo{journal}{Nat.~Photonics} \textbf{\bibinfo{volume}{10}},
  \bibinfo{pages}{258} (\bibinfo{year}{2016}).
\bibitem [{\citenamefont {Huntemann}\ \emph {et~al.}(2016)\citenamefont
		{Huntemann}, \citenamefont {Sanner}, \citenamefont {Lipphardt}, \citenamefont
		{Tamm},\ and\ \citenamefont {Peik}}]{huntemann2016single}%
	\BibitemOpen
	\bibfield  {author} {\bibinfo {author} {\bibfnamefont {N.}~\bibnamefont
			{Huntemann}}, \bibinfo {author} {\bibfnamefont {C.}~\bibnamefont {Sanner}},
		\bibinfo {author} {\bibfnamefont {B.}~\bibnamefont {Lipphardt}}, \bibinfo
		{author} {\bibfnamefont {C.}~\bibnamefont {Tamm}}, \ and\ \bibinfo {author}
		{\bibfnamefont {E.}~\bibnamefont {Peik}},\ }\href@noop {} {\bibfield
		{journal} {\bibinfo  {journal} {Phys.~Rev.~Lett.}\ }\textbf {\bibinfo
			{volume} {116}},\ \bibinfo {pages} {063001} (\bibinfo {year}
		{2016})}\BibitemShut {NoStop}%
	\bibitem [{\citenamefont {Stadnik}\ and\ \citenamefont
		{Flambaum}(2016)}]{stadnik2016enhanced}%
	\BibitemOpen
	\bibfield  {author} {\bibinfo {author} {\bibfnamefont {Y.}~\bibnamefont
			{Stadnik}}\ and\ \bibinfo {author} {\bibfnamefont {V.}~\bibnamefont
			{Flambaum}},\ }\href@noop {} {\bibfield  {journal} {\bibinfo  {journal}
			{Phys.~Rev.~A}\ }\textbf {\bibinfo {volume} {93}},\ \bibinfo {pages}
		{063630} (\bibinfo {year} {2016})}\BibitemShut {NoStop}%
	\bibitem [{\citenamefont {Derevianko}\ and\ \citenamefont
		{Pospelov}(2014)}]{derevianko2014hunting}%
	\BibitemOpen
	\bibfield  {author} {\bibinfo {author} {\bibfnamefont {A.}~\bibnamefont
			{Derevianko}}\ and\ \bibinfo {author} {\bibfnamefont {M.}~\bibnamefont
			{Pospelov}},\ }\href@noop {} {\bibfield  {journal} {\bibinfo  {journal}
			{Nat.~Physics}\ }\textbf {\bibinfo {volume} {10}},\ \bibinfo {pages} {933}
		(\bibinfo {year} {2014})}\BibitemShut {NoStop}%
	\bibitem [{\citenamefont {Safronova}\ \emph {et~al.}(2018)\citenamefont
		{Safronova}, \citenamefont {Budker}, \citenamefont {DeMille}, \citenamefont
		{Kimball}, \citenamefont {Derevianko},\ and\ \citenamefont
		{Clark}}]{safronova2018search}%
	\BibitemOpen
	\bibfield  {author} {\bibinfo {author} {\bibfnamefont {M.}~\bibnamefont
			{Safronova}}, \bibinfo {author} {\bibfnamefont {D.}~\bibnamefont {Budker}},
		\bibinfo {author} {\bibfnamefont {D.}~\bibnamefont {DeMille}}, \bibinfo
		{author} {\bibfnamefont {D.~F.~J.}\ \bibnamefont {Kimball}}, \bibinfo
		{author} {\bibfnamefont {A.}~\bibnamefont {Derevianko}}, \ and\ \bibinfo
		{author} {\bibfnamefont {C.~W.}\ \bibnamefont {Clark}},\ }\href@noop {}
	{\bibfield  {journal} {\bibinfo  {journal} {Rev.~Mod.~Phys.}\
		}\textbf {\bibinfo {volume} {90}},\ \bibinfo {pages} {025008} (\bibinfo
		{year} {2018})}\BibitemShut {NoStop}%
	\bibitem [{\citenamefont {Riehle}(2015)}]{riehle2015towards}%
	\BibitemOpen
	\bibfield  {author} {\bibinfo {author} {\bibfnamefont {F.}~\bibnamefont
			{Riehle}},\ }\href@noop {} {\bibfield  {journal} {\bibinfo  {journal}
			{Compt.~Rend.~Phys.}\ }\textbf {\bibinfo {volume} {16}},\ \bibinfo
		{pages} {506} (\bibinfo {year} {2015})}\BibitemShut {NoStop}%
	\bibitem [{\citenamefont {Ashby}\ \emph {et~al.}(2018)\citenamefont {Ashby},
		\citenamefont {Parker},\ and\ \citenamefont {Patla}}]{ashby2018null}%
	\BibitemOpen
	\bibfield  {author} {\bibinfo {author} {\bibfnamefont {N.}~\bibnamefont
			{Ashby}}, \bibinfo {author} {\bibfnamefont {T.~E.}\ \bibnamefont {Parker}}, \
		and\ \bibinfo {author} {\bibfnamefont {B.~R.}\ \bibnamefont {Patla}},\
	}\href@noop {} {\bibfield  {journal} {\bibinfo  {journal} {Nat.~Physics}\ \textbf {\bibinfo {volume} {14}},
			\ \bibinfo {pages} {822}} (\bibinfo {year} {2018})}\BibitemShut {NoStop}%
	\bibitem [{\citenamefont {Georgescu}(2017)}]{georgescu2017special}%
	\BibitemOpen
	\bibfield  {author} {\bibinfo {author} {\bibfnamefont {I.}~\bibnamefont
			{Georgescu}},\ }\href@noop {} {\bibfield  {journal} {\bibinfo  {journal}
			{Nat.~Physics}\ }\textbf {\bibinfo {volume} {13}},\ \bibinfo {pages} {529}
		(\bibinfo {year} {2017})}\BibitemShut {NoStop}%
	\bibitem [{\citenamefont {Chou}\ \emph {et~al.}(2010)\citenamefont {Chou},
		\citenamefont {Hume}, \citenamefont {Rosenband},\ and\ \citenamefont
		{Wineland}}]{chou2010optical}%
	\BibitemOpen
	\bibfield  {author} {\bibinfo {author} {\bibfnamefont {C.-W.}\ \bibnamefont
			{Chou}}, \bibinfo {author} {\bibfnamefont {D.}~\bibnamefont {Hume}}, \bibinfo
		{author} {\bibfnamefont {T.}~\bibnamefont {Rosenband}}, \ and\ \bibinfo
		{author} {\bibfnamefont {D.}~\bibnamefont {Wineland}},\ }\href@noop {}
	{\bibfield  {journal} {\bibinfo  {journal} {Science}\ }\textbf {\bibinfo
			{volume} {329}},\ \bibinfo {pages} {1630} (\bibinfo {year}
		{2010})}\BibitemShut {NoStop}%
	\bibitem [{\citenamefont {Abbott}\ \emph {et~al.}(2016)\citenamefont {Abbott},
		\citenamefont {Abbott}, \citenamefont {Abbott}, \citenamefont {Abernathy},
		\citenamefont {Acernese}, \citenamefont {Ackley}, \citenamefont {Adams},
		\citenamefont {Adams}, \citenamefont {Addesso}, \citenamefont {Adhikari}
		\emph {et~al.}}]{abbott2016observation}%
	\BibitemOpen
	\bibfield  {author} {\bibinfo {author} {\bibfnamefont {B.~P.}\ \bibnamefont
			{Abbott}}, \bibinfo {author} {\bibfnamefont {R.}~\bibnamefont {Abbott}},
		\bibinfo {author} {\bibfnamefont {T.~D.}\ \bibnamefont {Abbott}}, \bibinfo
		{author} {\bibfnamefont {M.~R.}\ \bibnamefont {Abernathy}}, \bibinfo {author}
		{\bibfnamefont {F.}~\bibnamefont {Acernese}}, \bibinfo {author}
		{\bibfnamefont {K.}~\bibnamefont {Ackley}}, \bibinfo {author} {\bibfnamefont
			{C.}~\bibnamefont {Adams}}, \bibinfo {author} {\bibfnamefont
			{T.}~\bibnamefont {Adams}}, \bibinfo {author} {\bibfnamefont
			{P.}~\bibnamefont {Addesso}}, \bibinfo {author} {\bibnamefont {Adhikari}},
		\emph {et~al.} (\bibinfo {collaboration} {LIGO Scientific Collaboration and
			Virgo Collaboration}),\ }\href {\doibase 10.1103/PhysRevLett.116.061102}
	{\bibfield  {journal} {\bibinfo  {journal} {Phys.~Rev.~Lett.}\ }\textbf
		{\bibinfo {volume} {116}},\ \bibinfo {pages} {061102} (\bibinfo {year}
		{2016})}\BibitemShut {NoStop}%
	\bibitem [{\citenamefont {Acernese}\ \emph {et~al.}(2014)\citenamefont
		{Acernese}, \citenamefont {Agathos}, \citenamefont {Agatsuma}, \citenamefont
		{Aisa}, \citenamefont {Allemandou}, \citenamefont {Allocca}, \citenamefont
		{Amarni}, \citenamefont {Astone}, \citenamefont {Balestri}, \citenamefont
		{Ballardin} \emph {et~al.}}]{acernese2014advanced}%
	\BibitemOpen
	\bibfield  {author} {\bibinfo {author} {\bibfnamefont {F.}~\bibnamefont
			{Acernese}}, \bibinfo {author} {\bibfnamefont {M.}~\bibnamefont {Agathos}},
		\bibinfo {author} {\bibfnamefont {K.}~\bibnamefont {Agatsuma}}, \bibinfo
		{author} {\bibfnamefont {D.}~\bibnamefont {Aisa}}, \bibinfo {author}
		{\bibfnamefont {N.}~\bibnamefont {Allemandou}}, \bibinfo {author}
		{\bibfnamefont {A.}~\bibnamefont {Allocca}}, \bibinfo {author} {\bibfnamefont
			{J.}~\bibnamefont {Amarni}}, \bibinfo {author} {\bibfnamefont
			{P.}~\bibnamefont {Astone}}, \bibinfo {author} {\bibfnamefont
			{G.}~\bibnamefont {Balestri}}, \bibinfo {author} {\bibfnamefont
			{G.}~\bibnamefont {Ballardin}},  \emph {et~al.},\ }\href {\doibase
		10.1088/0264-9381/32/2/024001} {\bibfield  {journal} {\bibinfo  {journal}
			{Class.~Quant.~Grav.}\ }\textbf {\bibinfo {volume} {32}},\ \bibinfo
		{pages} {024001} (\bibinfo {year} {2014})}\BibitemShut {NoStop}%
	\bibitem[{\citenamefont{Canuel et~al.}(2018)\citenamefont{Canuel, Bertoldi,
  Amand, Pozzo~di Borgo, Chantrait, Danquigny, Dovale~Álvarez, Fang, Freise,
  Geiger}, \emph {et~al.}}]{Canuel2018}
\bibinfo{author}{\bibfnamefont{B.}~\bibnamefont{Canuel}},
  \bibinfo{author}{\bibfnamefont{A.}~\bibnamefont{Bertoldi}},
  \bibinfo{author}{\bibfnamefont{L.}~\bibnamefont{Amand}},
  \bibinfo{author}{\bibfnamefont{E.}~\bibnamefont{Pozzo~di Borgo}},
  \bibinfo{author}{\bibfnamefont{T.}~\bibnamefont{Chantrait}},
  \bibinfo{author}{\bibfnamefont{C.}~\bibnamefont{Danquigny}},
  \bibinfo{author}{\bibfnamefont{M.}~\bibnamefont{Dovale~Álvarez}},
  \bibinfo{author}{\bibfnamefont{B.}~\bibnamefont{Fang}},
  \bibinfo{author}{\bibfnamefont{A.}~\bibnamefont{Freise}},
  \bibinfo{author}{\bibfnamefont{R.}~\bibnamefont{Geiger}},
  \bibnamefont\emph {et~al.}, \bibinfo{journal}{Scient.~Rep.}
  \textbf{\bibinfo{volume}{8}}, \bibinfo{pages}{14064} (\bibinfo{year}{2018}),

	\bibitem [{\citenamefont {Zhang}\ \emph {et~al.}(2014)\citenamefont {Zhang},
		\citenamefont {Bishof}, \citenamefont {Bromley}, \citenamefont {Kraus},
		\citenamefont {Safronova}, \citenamefont {Zoller}, \citenamefont {Rey},\ and\
		\citenamefont {Ye}}]{zhang2014spectroscopic}%
	\BibitemOpen
	\bibfield  {author} {\bibinfo {author} {\bibfnamefont {X.}~\bibnamefont
			{Zhang}}, \bibinfo {author} {\bibfnamefont {M.}~\bibnamefont {Bishof}},
		\bibinfo {author} {\bibfnamefont {S.~L.}\ \bibnamefont {Bromley}}, \bibinfo
		{author} {\bibfnamefont {C.~V.}\ \bibnamefont {Kraus}}, \bibinfo {author}
		{\bibfnamefont {M.~S.}\ \bibnamefont {Safronova}}, \bibinfo {author}
		{\bibfnamefont {P.}~\bibnamefont {Zoller}}, \bibinfo {author} {\bibfnamefont
			{A.~M.}\ \bibnamefont {Rey}}, \ and\ \bibinfo {author} {\bibfnamefont
			{J.}~\bibnamefont {Ye}},\ }\href@noop {} {\bibfield  {journal} {\bibinfo
			{journal} {Science}\ }\textbf {\bibinfo {volume} {345}},\ \bibinfo {pages}
		{1467} (\bibinfo {year} {2014})}\BibitemShut {NoStop}%
	\bibitem [{\citenamefont {Schmidt-Kaler}\ \emph {et~al.}(2003)\citenamefont
		{Schmidt-Kaler}, \citenamefont {H{\"a}ffner}, \citenamefont {Riebe},
		\citenamefont {Gulde}, \citenamefont {Lancaster}, \citenamefont {Deuschle},
		\citenamefont {Becher}, \citenamefont {Roos}, \citenamefont {Eschner},\ and\
		\citenamefont {Blatt}}]{schmidt2003realization}%
	\BibitemOpen
	\bibfield  {author} {\bibinfo {author} {\bibfnamefont {F.}~\bibnamefont
			{Schmidt-Kaler}}, \bibinfo {author} {\bibfnamefont {H.}~\bibnamefont
			{H{\"a}ffner}}, \bibinfo {author} {\bibfnamefont {M.}~\bibnamefont {Riebe}},
		\bibinfo {author} {\bibfnamefont {S.}~\bibnamefont {Gulde}}, \bibinfo
		{author} {\bibfnamefont {G.~P.}\ \bibnamefont {Lancaster}}, \bibinfo {author}
		{\bibfnamefont {T.}~\bibnamefont {Deuschle}}, \bibinfo {author}
		{\bibfnamefont {C.}~\bibnamefont {Becher}}, \bibinfo {author} {\bibfnamefont
			{C.~F.}\ \bibnamefont {Roos}}, \bibinfo {author} {\bibfnamefont
			{J.}~\bibnamefont {Eschner}}, \ and\ \bibinfo {author} {\bibfnamefont
			{R.}~\bibnamefont {Blatt}},\ }\href@noop {} {\bibfield  {journal} {\bibinfo
			{journal} {Nature}\ }\textbf {\bibinfo {volume} {422}},\ \bibinfo {pages}
		{408} (\bibinfo {year} {2003})}\BibitemShut {NoStop}%
	\bibitem [{\citenamefont {Leopardi}\ \emph {et~al.}(2017)\citenamefont
		{Leopardi}, \citenamefont {Davila-Rodriguez}, \citenamefont {Quinlan},
		\citenamefont {Olson}, \citenamefont {Sherman}, \citenamefont {Diddams},\
		and\ \citenamefont {Fortier}}]{leopardi2017single}%
	\BibitemOpen
	\bibfield  {author} {\bibinfo {author} {\bibfnamefont {H.}~\bibnamefont
			{Leopardi}}, \bibinfo {author} {\bibfnamefont {J.}~\bibnamefont
			{Davila-Rodriguez}}, \bibinfo {author} {\bibfnamefont {F.}~\bibnamefont
			{Quinlan}}, \bibinfo {author} {\bibfnamefont {J.}~\bibnamefont {Olson}},
		\bibinfo {author} {\bibfnamefont {J.~A.}\ \bibnamefont {Sherman}}, \bibinfo
		{author} {\bibfnamefont {S.~A.}\ \bibnamefont {Diddams}}, \ and\ \bibinfo
		{author} {\bibfnamefont {T.~M.}\ \bibnamefont {Fortier}},\ }\href@noop {}
	{\bibfield  {journal} {\bibinfo  {journal} {Optica}\ }\textbf {\bibinfo
			{volume} {4}},\ \bibinfo {pages} {879} (\bibinfo {year} {2017})}\BibitemShut
	{NoStop}%
    \bibitem[{\citenamefont{Xie et~al.}(2016)\citenamefont{Xie, Bouchand, Nicolodi,
  Giunta, Hänsel, Lezius, Joshi, Datta, Alexandre, Lours et~al.}}]{Xie2016}
\bibinfo{author}{\bibfnamefont{X.}~\bibnamefont{Xie}},
  \bibinfo{author}{\bibfnamefont{R.}~\bibnamefont{Bouchand}},
  \bibinfo{author}{\bibfnamefont{D.}~\bibnamefont{Nicolodi}},
  \bibinfo{author}{\bibfnamefont{M.}~\bibnamefont{Giunta}},
  \bibinfo{author}{\bibfnamefont{W.}~\bibnamefont{Hänsel}},
  \bibinfo{author}{\bibfnamefont{M.}~\bibnamefont{Lezius}},
  \bibinfo{author}{\bibfnamefont{A.}~\bibnamefont{Joshi}},
  \bibinfo{author}{\bibfnamefont{S.}~\bibnamefont{Datta}},
  \bibinfo{author}{\bibfnamefont{C.}~\bibnamefont{Alexandre}},
  \bibinfo{author}{\bibfnamefont{M.}~\bibnamefont{Lours}},
  \bibnamefont{\emph {et~al.}}, \bibinfo{journal}{Nat.~Photonics}
  \textbf{\bibinfo{volume}{11}}, \bibinfo{pages}{44} (\bibinfo{year}{2016}).
	
    \bibitem [{\citenamefont {H{\"a}fner}\ \emph {et~al.}(2015)\citenamefont
		{H{\"a}fner}, \citenamefont {Falke}, \citenamefont {Grebing}, \citenamefont
		{Vogt}, \citenamefont {Legero}, \citenamefont {Merimaa}, \citenamefont
		{Lisdat},\ and\ \citenamefont {Sterr}}]{hafner20158}%
	\BibitemOpen
	\bibfield  {author} {\bibinfo {author} {\bibfnamefont {S.}~\bibnamefont
			{H{\"a}fner}}, \bibinfo {author} {\bibfnamefont {S.}~\bibnamefont {Falke}},
		\bibinfo {author} {\bibfnamefont {C.}~\bibnamefont {Grebing}}, \bibinfo
		{author} {\bibfnamefont {S.}~\bibnamefont {Vogt}}, \bibinfo {author}
		{\bibfnamefont {T.}~\bibnamefont {Legero}}, \bibinfo {author} {\bibfnamefont
			{M.}~\bibnamefont {Merimaa}}, \bibinfo {author} {\bibfnamefont
			{C.}~\bibnamefont {Lisdat}}, \ and\ \bibinfo {author} {\bibfnamefont
			{U.}~\bibnamefont {Sterr}},\ }\href@noop {} {\bibfield  {journal} {\bibinfo
			{journal} {Opt.~Lett.}\ }\textbf {\bibinfo {volume} {40}},\ \bibinfo
		{pages} {2112} (\bibinfo {year} {2015})}\BibitemShut {NoStop}%
	\bibitem [{\citenamefont {Matei}\ \emph {et~al.}(2017)\citenamefont {Matei},
		\citenamefont {Legero}, \citenamefont {H\"afner}, \citenamefont {Grebing},
		\citenamefont {Weyrich}, \citenamefont {Zhang}, \citenamefont {Sonderhouse},
		\citenamefont {Robinson}, \citenamefont {Ye} \emph {et~al.}}]{matei20171}%
	\BibitemOpen
	\bibfield  {author} {\bibinfo {author} {\bibfnamefont {D.~G.}\ \bibnamefont
			{Matei}}, \bibinfo {author} {\bibfnamefont {T.}~\bibnamefont {Legero}},
		\bibinfo {author} {\bibfnamefont {S.}~\bibnamefont {H\"afner}}, \bibinfo
		{author} {\bibfnamefont {C.}~\bibnamefont {Grebing}}, \bibinfo {author}
		{\bibfnamefont {R.}~\bibnamefont {Weyrich}}, \bibinfo {author} {\bibfnamefont
			{W.}~\bibnamefont {Zhang}}, \bibinfo {author} {\bibfnamefont
			{L.}~\bibnamefont {Sonderhouse}}, \bibinfo {author} {\bibfnamefont {J.~M.}\
			\bibnamefont {Robinson}}, \bibinfo {author} {\bibfnamefont {J.}~\bibnamefont
			{Ye}},  \emph {et~al.},\ }\href {\doibase 10.1103/PhysRevLett.118.263202}
	{\bibfield  {journal} {\bibinfo  {journal} {Phys.~Rev.~Lett.}\ }\textbf
		{\bibinfo {volume} {118}},\ \bibinfo {pages} {263202} (\bibinfo {year}
		{2017})}\BibitemShut {NoStop}%
	\bibitem [{\citenamefont {Robinson}\ \emph {et~al.}(2019)\citenamefont
		{Robinson}, \citenamefont {Oelker}, \citenamefont {Milner}, \citenamefont
		{Zhang}, \citenamefont {Legero}, \citenamefont {Matei}, \citenamefont
		{Riehle}, \citenamefont {Sterr},\ and\ \citenamefont
		{Ye}}]{robinson2019crystalline}%
	\BibitemOpen
	\bibfield  {author} {\bibinfo {author} {\bibfnamefont {J.~M.}\ \bibnamefont
			{Robinson}}, \bibinfo {author} {\bibfnamefont {E.}~\bibnamefont {Oelker}},
		\bibinfo {author} {\bibfnamefont {W.~R.}\ \bibnamefont {Milner}}, \bibinfo
		{author} {\bibfnamefont {W.}~\bibnamefont {Zhang}}, \bibinfo {author}
		{\bibfnamefont {T.}~\bibnamefont {Legero}}, \bibinfo {author} {\bibfnamefont
			{D.~G.}\ \bibnamefont {Matei}}, \bibinfo {author} {\bibfnamefont
			{F.}~\bibnamefont {Riehle}}, \bibinfo {author} {\bibfnamefont
			{U.}~\bibnamefont {Sterr}}, \ and\ \bibinfo {author} {\bibfnamefont
			{J.}~\bibnamefont {Ye}},\ }\href@noop {} {\bibfield  {journal} {\bibinfo
			{journal} {Optica}\ }\textbf {\bibinfo {volume} {6}},\ \bibinfo {pages} {240}
		(\bibinfo {year} {2019})}\BibitemShut {NoStop}%
	\bibitem [{\citenamefont {Jiang}\ \emph {et~al.}(2011)\citenamefont {Jiang},
		\citenamefont {Ludlow}, \citenamefont {Lemke}, \citenamefont {Fox},
		\citenamefont {Sherman}, \citenamefont {Ma},\ and\ \citenamefont
		{Oates}}]{jiang2011making}%
	\BibitemOpen
	\bibfield  {author} {\bibinfo {author} {\bibfnamefont {Y.}~\bibnamefont
			{Jiang}}, \bibinfo {author} {\bibfnamefont {A.}~\bibnamefont {Ludlow}},
		\bibinfo {author} {\bibfnamefont {N.~D.}\ \bibnamefont {Lemke}}, \bibinfo
		{author} {\bibfnamefont {R.~W.}\ \bibnamefont {Fox}}, \bibinfo {author}
		{\bibfnamefont {J.~A.}\ \bibnamefont {Sherman}}, \bibinfo {author}
		{\bibfnamefont {L.-S.}\ \bibnamefont {Ma}}, \ and\ \bibinfo {author}
		{\bibfnamefont {C.~W.}\ \bibnamefont {Oates}},\ }\href@noop {} {\bibfield
		{journal} {\bibinfo  {journal} {Nat.~Photonics}\ }\textbf {\bibinfo
			{volume} {5}},\ \bibinfo {pages} {158} (\bibinfo {year} {2011})}\BibitemShut
	{NoStop}%
	\bibitem [{\citenamefont {Numata}\ \emph {et~al.}(2004)\citenamefont {Numata},
		\citenamefont {Kemery},\ and\ \citenamefont {Camp}}]{numata2004thermal}%
	\BibitemOpen
	\bibfield  {author} {\bibinfo {author} {\bibfnamefont {K.}~\bibnamefont
			{Numata}}, \bibinfo {author} {\bibfnamefont {A.}~\bibnamefont {Kemery}}, \
		and\ \bibinfo {author} {\bibfnamefont {J.}~\bibnamefont {Camp}},\ }\href@noop
	{} {\bibfield  {journal} {\bibinfo  {journal} {Phys.~Rev.~Lett.}\
		}\textbf {\bibinfo {volume} {93}},\ \bibinfo {pages} {250602} (\bibinfo
		{year} {2004})}\BibitemShut {NoStop}%
	\bibitem [{\citenamefont {Norcia}\ \emph {et~al.}(2018)\citenamefont {Norcia},
		\citenamefont {Cline}, \citenamefont {Muniz}, \citenamefont {Robinson},
		\citenamefont {Hutson}, \citenamefont {Goban}, \citenamefont {Marti},
		\citenamefont {Ye},\ and\ \citenamefont {Thompson}}]{norcia2018frequency}%
	\BibitemOpen
	\bibfield  {author} {\bibinfo {author} {\bibfnamefont {M.~A.}\ \bibnamefont
			{Norcia}}, \bibinfo {author} {\bibfnamefont {J.~R.}\ \bibnamefont {Cline}},
		\bibinfo {author} {\bibfnamefont {J.~A.}\ \bibnamefont {Muniz}}, \bibinfo
		{author} {\bibfnamefont {J.~M.}\ \bibnamefont {Robinson}}, \bibinfo {author}
		{\bibfnamefont {R.~B.}\ \bibnamefont {Hutson}}, \bibinfo {author}
		{\bibfnamefont {A.}~\bibnamefont {Goban}}, \bibinfo {author} {\bibfnamefont
			{G.~E.}\ \bibnamefont {Marti}}, \bibinfo {author} {\bibfnamefont
			{J.}~\bibnamefont {Ye}}, \ and\ \bibinfo {author} {\bibfnamefont {J.~K.}\
			\bibnamefont {Thompson}},\ }\href@noop {} {\bibfield  {journal} {\bibinfo
			{journal} {Phys.~Rev.~X}\ }\textbf {\bibinfo {volume} {8}},\ \bibinfo
		{pages} {021036} (\bibinfo {year} {2018})}\BibitemShut {NoStop}%
	\bibitem [{\citenamefont {Sch{\"a}ffer}\ \emph {et~al.}(2017)\citenamefont
		{Sch{\"a}ffer}, \citenamefont {Christensen}, \citenamefont {Rathmann},
		\citenamefont {Appel}, \citenamefont {Henriksen},\ and\ \citenamefont
		{Thomsen}}]{schaffer2017towards}%
	\BibitemOpen
	\bibfield  {author} {\bibinfo {author} {\bibfnamefont {S.~A.}\ \bibnamefont
			{Sch{\"a}ffer}}, \bibinfo {author} {\bibfnamefont {B.~T.~R.}\ \bibnamefont
			{Christensen}}, \bibinfo {author} {\bibfnamefont {S.~M.}\ \bibnamefont
			{Rathmann}}, \bibinfo {author} {\bibfnamefont {M.~H.}\ \bibnamefont {Appel}},
		\bibinfo {author} {\bibfnamefont {M.~R.}\ \bibnamefont {Henriksen}}, \ and\
		\bibinfo {author} {\bibfnamefont {J.~W.}\ \bibnamefont {Thomsen}},\ }in\
	\href@noop {} {\emph {\bibinfo {booktitle} {J.~Phys.: Conf.~Series}}},\ Vol.\ \bibinfo {volume} {810}\ (\bibinfo {organization} {IOP
		Publishing},\ \bibinfo {year} {2017})\ p.\ \bibinfo {pages}
	{012002}\BibitemShut {NoStop}%
	\bibitem [{\citenamefont {Wang}\ and\ \citenamefont
		{Chen}(2012)}]{wang2012superradiant}%
	\BibitemOpen
	\bibfield  {author} {\bibinfo {author} {\bibfnamefont {Y.-F.}\ \bibnamefont
			{Wang}}\ and\ \bibinfo {author} {\bibfnamefont {J.-B.}\ \bibnamefont
			{Chen}},\ }\href@noop {} {\bibfield  {journal} {\bibinfo  {journal} {Chin.~Phys.~Lett.}\ }\textbf {\bibinfo {volume} {29}},\ \bibinfo {pages}
		{73202} (\bibinfo {year} {2012})}\BibitemShut {NoStop}%
	\bibitem [{\citenamefont {Cook}\ \emph {et~al.}(2015)\citenamefont {Cook},
		\citenamefont {Rosenband},\ and\ \citenamefont {Leibrandt}}]{cook2015laser}%
	\BibitemOpen
	\bibfield  {author} {\bibinfo {author} {\bibfnamefont {S.}~\bibnamefont
			{Cook}}, \bibinfo {author} {\bibfnamefont {T.}~\bibnamefont {Rosenband}}, \
		and\ \bibinfo {author} {\bibfnamefont {D.~R.}\ \bibnamefont {Leibrandt}},\
	}\href@noop {} {\bibfield  {journal} {\bibinfo  {journal} {Phys.~Rev.~
				Lett.}\ }\textbf {\bibinfo {volume} {114}},\ \bibinfo {pages} {253902}
		(\bibinfo {year} {2015})}\BibitemShut {NoStop}%
	\bibitem [{\citenamefont {Bergquist}\ \emph {et~al.}(1977)\citenamefont
		{Bergquist}, \citenamefont {Lee},\ and\ \citenamefont
		{Hall}}]{bergquist1977saturated}%
	\BibitemOpen
	\bibfield  {author} {\bibinfo {author} {\bibfnamefont {J.}~\bibnamefont
			{Bergquist}}, \bibinfo {author} {\bibfnamefont {S.}~\bibnamefont {Lee}}, \
		and\ \bibinfo {author} {\bibfnamefont {J.}~\bibnamefont {Hall}},\ }\href@noop
	{} {\bibfield  {journal} {\bibinfo  {journal} {Phys.~Rev.~Lett.}\
		}\textbf {\bibinfo {volume} {38}},\ \bibinfo {pages} {159} (\bibinfo {year}
		{1977})}\BibitemShut {NoStop}%
	\bibitem [{\citenamefont {Barger}\ \emph {et~al.}(1979)\citenamefont {Barger},
		\citenamefont {Bergquist}, \citenamefont {English},\ and\ \citenamefont
		{Glaze}}]{barger1979resolution}%
	\BibitemOpen
	\bibfield  {author} {\bibinfo {author} {\bibfnamefont {R.}~\bibnamefont
			{Barger}}, \bibinfo {author} {\bibfnamefont {J.}~\bibnamefont {Bergquist}},
		\bibinfo {author} {\bibfnamefont {T.}~\bibnamefont {English}}, \ and\
		\bibinfo {author} {\bibfnamefont {D.}~\bibnamefont {Glaze}},\ }\href@noop {}
	{\bibfield  {journal} {\bibinfo  {journal} {Appl.~Phys.~Lett.}\
		}\textbf {\bibinfo {volume} {34}},\ \bibinfo {pages} {850} (\bibinfo {year}
		{1979})}\BibitemShut {NoStop}%
	\bibitem [{\citenamefont {Bord{\'e}}(1989)}]{borde1989atomic}%
	\BibitemOpen
	\bibfield  {author} {\bibinfo {author} {\bibfnamefont {C.~J.}\ \bibnamefont
			{Bord{\'e}}},\ }\href@noop {} {\bibfield  {journal} {\bibinfo  {journal}
			{Phys.~Lett.~A}\ }\textbf {\bibinfo {volume} {140}},\ \bibinfo {pages}
		{10} (\bibinfo {year} {1989})}\BibitemShut {NoStop}%
	\bibitem [{\citenamefont {Kovachy}\ \emph {et~al.}(2015)\citenamefont
		{Kovachy}, \citenamefont {Asenbaum}, \citenamefont {Overstreet},
		\citenamefont {Donnelly}, \citenamefont {Dickerson}, \citenamefont
		{Sugarbaker}, \citenamefont {Hogan},\ and\ \citenamefont
		{Kasevich}}]{kovachy2015quantum}%
	\BibitemOpen
	\bibfield  {author} {\bibinfo {author} {\bibfnamefont {T.}~\bibnamefont
			{Kovachy}}, \bibinfo {author} {\bibfnamefont {P.}~\bibnamefont {Asenbaum}},
		\bibinfo {author} {\bibfnamefont {C.}~\bibnamefont {Overstreet}}, \bibinfo
		{author} {\bibfnamefont {C.}~\bibnamefont {Donnelly}}, \bibinfo {author}
		{\bibfnamefont {S.}~\bibnamefont {Dickerson}}, \bibinfo {author}
		{\bibfnamefont {A.}~\bibnamefont {Sugarbaker}}, \bibinfo {author}
		{\bibfnamefont {J.}~\bibnamefont {Hogan}}, \ and\ \bibinfo {author}
		{\bibfnamefont {M.}~\bibnamefont {Kasevich}},\ }\href@noop {} {\bibfield
		{journal} {\bibinfo  {journal} {Nature}\ }\textbf {\bibinfo {volume} {528}},\
		\bibinfo {pages} {530} (\bibinfo {year} {2015})}\BibitemShut {NoStop}%
	\bibitem [{\citenamefont {Hamilton}\ \emph {et~al.}(2015)\citenamefont
		{Hamilton}, \citenamefont {Jaffe}, \citenamefont {Haslinger}, \citenamefont
		{Simmons}, \citenamefont {M{\"u}ller},\ and\ \citenamefont
		{Khoury}}]{hamilton2015atom}%
	\BibitemOpen
	\bibfield  {author} {\bibinfo {author} {\bibfnamefont {P.}~\bibnamefont
			{Hamilton}}, \bibinfo {author} {\bibfnamefont {M.}~\bibnamefont {Jaffe}},
		\bibinfo {author} {\bibfnamefont {P.}~\bibnamefont {Haslinger}}, \bibinfo
		{author} {\bibfnamefont {Q.}~\bibnamefont {Simmons}}, \bibinfo {author}
		{\bibfnamefont {H.}~\bibnamefont {M{\"u}ller}}, \ and\ \bibinfo {author}
		{\bibfnamefont {J.}~\bibnamefont {Khoury}},\ }\href@noop {} {\bibfield
		{journal} {\bibinfo  {journal} {Science}\ }\textbf {\bibinfo {volume}
			{349}},\ \bibinfo {pages} {849} (\bibinfo {year} {2015})}\BibitemShut
	{NoStop}%
	\bibitem [{\citenamefont {Dubetsky}\ \emph {et~al.}(2016)\citenamefont
		{Dubetsky}, \citenamefont {Libby},\ and\ \citenamefont
		{Berman}}]{dubetsky2016atom}%
	\BibitemOpen
	\bibfield  {author} {\bibinfo {author} {\bibfnamefont {B.}~\bibnamefont
			{Dubetsky}}, \bibinfo {author} {\bibfnamefont {S.~B.}\ \bibnamefont {Libby}},
		\ and\ \bibinfo {author} {\bibfnamefont {P.}~\bibnamefont {Berman}},\
	}\href@noop {} {\bibfield  {journal} {\bibinfo  {journal} {Atoms}\ }\textbf
		{\bibinfo {volume} {4}},\ \bibinfo {pages} {14} (\bibinfo {year}
		{2016})}\BibitemShut {NoStop}%
	\bibitem [{\citenamefont {Keupp}\ \emph {et~al.}(2005)\citenamefont {Keupp},
  \citenamefont {Douillet}, \citenamefont {Mehlst{\"a}ubler}, \citenamefont
  {Rehbein}, \citenamefont {Rasel},\ and\ \citenamefont {Ertmer}}]{Keupp2005}%
  \BibitemOpen
  \bibfield  {author} {\bibinfo {author} {\bibfnamefont {J.}~\bibnamefont
  {Keupp}}, \bibinfo {author} {\bibfnamefont {A.}~\bibnamefont {Douillet}},
  \bibinfo {author} {\bibfnamefont {T.~E.}\ \bibnamefont {Mehlst{\"a}ubler}},
  \bibinfo {author} {\bibfnamefont {N.}~\bibnamefont {Rehbein}}, \bibinfo
  {author} {\bibfnamefont {E.~M.}\ \bibnamefont {Rasel}}, \ and\ \bibinfo
  {author} {\bibfnamefont {W.}~\bibnamefont {Ertmer}},\ }\href {\doibase
  10.1140/epjd/e2005-00302-7} {\bibfield  {journal} {\bibinfo  {journal} {Eur.~Phys.~J.~D.}\
  }\textbf {\bibinfo {volume} {36}},\ \bibinfo {pages} {289} (\bibinfo {year}
  {2005})}\BibitemShut {NoStop}%
    \bibitem [{\citenamefont {Bord{\'e}}\ \emph {et~al.}(1984)\citenamefont
		{Bord{\'e}}, \citenamefont {Salomon}, \citenamefont {Avrillier},
		\citenamefont {Van~Lerberghe}, \citenamefont {Br{\'e}ant}, \citenamefont
		{Bassi},\ and\ \citenamefont {Scoles}}]{borde1984optical}%
	\BibitemOpen
	\bibfield  {author} {\bibinfo {author} {\bibfnamefont {C.~J.}\ \bibnamefont
			{Bord{\'e}}}, \bibinfo {author} {\bibfnamefont {C.}~\bibnamefont {Salomon}},
		\bibinfo {author} {\bibfnamefont {S.}~\bibnamefont {Avrillier}}, \bibinfo
		{author} {\bibfnamefont {A.}~\bibnamefont {Van~Lerberghe}}, \bibinfo {author}
		{\bibfnamefont {C.}~\bibnamefont {Br{\'e}ant}}, \bibinfo {author}
		{\bibfnamefont {D.}~\bibnamefont {Bassi}}, \ and\ \bibinfo {author}
		{\bibfnamefont {G.}~\bibnamefont {Scoles}},\ }\href@noop {} {\bibfield
		{journal} {\bibinfo  {journal} {Phys.~Rev.~A}\ }\textbf {\bibinfo
			{volume} {30}},\ \bibinfo {pages} {1836} (\bibinfo {year}
		{1984})}\BibitemShut {NoStop}%
	\bibitem [{\citenamefont {Barger}(1981)}]{barger1981influence}%
	\BibitemOpen
	\bibfield  {author} {\bibinfo {author} {\bibfnamefont {R.}~\bibnamefont
			{Barger}},\ }\href@noop {} {\bibfield  {journal} {\bibinfo  {journal} {Opt.~
				Lett.}\ }\textbf {\bibinfo {volume} {6}},\ \bibinfo {pages} {145} (\bibinfo
		{year} {1981})}\BibitemShut {NoStop}%
	\bibitem[{\citenamefont{Ito et~al.}(1994)\citenamefont{Ito, Ishikawa, and
  Morinaga}}]{Ito1994}
\bibinfo{author}{\bibfnamefont{N.}~\bibnamefont{Ito}},
  \bibinfo{author}{\bibfnamefont{J.}~\bibnamefont{Ishikawa}}, \bibnamefont{and}
  \bibinfo{author}{\bibfnamefont{A.}~\bibnamefont{Morinaga}},
  \bibinfo{journal}{Opt.~Commun.} \textbf{\bibinfo{volume}{109}},
  \bibinfo{pages}{414 } (\bibinfo{year}{1994}).
   \bibitem [{\citenamefont {Wilpers}\ \emph {et~al.}(2003)\citenamefont
  {Wilpers}, \citenamefont {Degenhardt}, \citenamefont {Binnewies},
  \citenamefont {Chernyshov}, \citenamefont {Riehle}, \citenamefont {Helmcke},\
  and\ \citenamefont {Sterr}}]{Wilpers2003}%
  \BibitemOpen
  \bibfield  {author} {\bibinfo {author} {\bibfnamefont {G.}~\bibnamefont
  {Wilpers}}, \bibinfo {author} {\bibfnamefont {C.}~\bibnamefont {Degenhardt}},
  \bibinfo {author} {\bibfnamefont {T.}~\bibnamefont {Binnewies}}, \bibinfo
  {author} {\bibfnamefont {A.}~\bibnamefont {Chernyshov}}, \bibinfo {author}
  {\bibfnamefont {F.}~\bibnamefont {Riehle}}, \bibinfo {author} {\bibfnamefont
  {J.}~\bibnamefont {Helmcke}}, \ and\ \bibinfo {author} {\bibfnamefont
  {U.}~\bibnamefont {Sterr}},\ }\href {\doibase 10.1007/s00340-003-1109-7}
  {\bibfield  {journal} {\bibinfo  {journal} {Appl.~Phys.~B}\ }\textbf
  {\bibinfo {volume} {76}},\ \bibinfo {pages} {149} (\bibinfo {year}
  {2003})}\BibitemShut {NoStop}%
  \bibitem [{\citenamefont {Webster}\ and\ \citenamefont
  {Gill}(2011)}]{Webster2011}%
  \BibitemOpen
  \bibfield  {author} {\bibinfo {author} {\bibfnamefont {S.}~\bibnamefont
  {Webster}}\ and\ \bibinfo {author} {\bibfnamefont {P.}~\bibnamefont {Gill}},\
  }\href {\doibase 10.1364/OL.36.003572} {\bibfield  {journal} {\bibinfo
  {journal} {Opt.~Lett.}\ }\textbf {\bibinfo {volume} {36}},\ \bibinfo {pages}
  {3572} (\bibinfo {year} {2011})}\BibitemShut {NoStop}%
  \bibitem [{\citenamefont {Leibrandt}\ \emph {et~al.}(2013)\citenamefont
  {Leibrandt}, \citenamefont {Bergquist},\ and\ \citenamefont
  {Rosenband}}]{Leibrandt2013}%
  \BibitemOpen
  \bibfield  {author} {\bibinfo {author} {\bibfnamefont {D.~R.}\ \bibnamefont
  {Leibrandt}}, \bibinfo {author} {\bibfnamefont {J.~C.}\ \bibnamefont
  {Bergquist}}, \ and\ \bibinfo {author} {\bibfnamefont {T.}~\bibnamefont
  {Rosenband}},\ }\href {\doibase 10.1103/PhysRevA.87.023829} {\bibfield
  {journal} {\bibinfo  {journal} {Phys.~Rev.~A}\ }\textbf {\bibinfo {volume}
  {87}},\ \bibinfo {pages} {023829} (\bibinfo {year} {2013})}\BibitemShut
  {NoStop}%
\bibitem [{\citenamefont {McFerran}\ and\ \citenamefont
  {Luiten}(2010)}]{McFerran2010}%
  \BibitemOpen
  \bibfield  {author} {\bibinfo {author} {\bibfnamefont {J.~J.}\ \bibnamefont
  {McFerran}}\ and\ \bibinfo {author} {\bibfnamefont {A.~N.}\ \bibnamefont
  {Luiten}},\ }\href {\doibase 10.1364/JOSAB.27.000277} {\bibfield  {journal}
  {\bibinfo  {journal} {J. Opt. Soc. Am. B}\ }\textbf {\bibinfo {volume}
  {27}},\ \bibinfo {pages} {277} (\bibinfo {year} {2010})}\BibitemShut
  {NoStop}%
  \bibitem [{\citenamefont {Shang}\ \emph {et~al.}(2017)\citenamefont {Shang},
  \citenamefont {Zhang}, \citenamefont {Zhang}, \citenamefont {Pan},
  \citenamefont {Chen},\ and\ \citenamefont {Chen}}]{Shang2017}%
  \BibitemOpen
  \bibfield  {author} {\bibinfo {author} {\bibfnamefont {H.}~\bibnamefont
  {Shang}}, \bibinfo {author} {\bibfnamefont {X.}~\bibnamefont {Zhang}},
  \bibinfo {author} {\bibfnamefont {S.}~\bibnamefont {Zhang}}, \bibinfo
  {author} {\bibfnamefont {D.}~\bibnamefont {Pan}}, \bibinfo {author}
  {\bibfnamefont {H.}~\bibnamefont {Chen}}, \ and\ \bibinfo {author}
  {\bibfnamefont {J.}~\bibnamefont {Chen}},\ }\href {\doibase
  10.1364/OE.25.030459} {\bibfield  {journal} {\bibinfo  {journal} {Opt.
  Express}\ }\textbf {\bibinfo {volume} {25}},\ \bibinfo {pages} {30459}
  (\bibinfo {year} {2017})}\BibitemShut {NoStop}%
\bibitem [{\citenamefont {Nagourney}\ \emph {et~al.}(1986)\citenamefont
		{Nagourney}, \citenamefont {Sandberg},\ and\ \citenamefont
		{Dehmelt}}]{nagourney1986shelved}%
	\BibitemOpen
	\bibfield  {author} {\bibinfo {author} {\bibfnamefont {W.}~\bibnamefont
			{Nagourney}}, \bibinfo {author} {\bibfnamefont {J.}~\bibnamefont {Sandberg}},
		\ and\ \bibinfo {author} {\bibfnamefont {H.}~\bibnamefont {Dehmelt}},\
	}\href@noop {} {\bibfield  {journal} {\bibinfo  {journal} {Phys.~Rev.~
				Lett.}\ }\textbf {\bibinfo {volume} {56}},\ \bibinfo {pages} {2797}
		(\bibinfo {year} {1986})}\BibitemShut {NoStop}%
	\bibitem [{\citenamefont {McGrew}\ \emph {et~al.}(2018)\citenamefont {McGrew},
		\citenamefont {Zhang}, \citenamefont {Fasano}, \citenamefont {Sch{\"a}ffer},
		\citenamefont {Beloy}, \citenamefont {Nicolodi}, \citenamefont {Brown},
		\citenamefont {Hinkley}, \citenamefont {Milani}, \citenamefont {Schioppo}
		\emph {et~al.}}]{mcgrew2018atomic}%
	\BibitemOpen
	\bibfield  {author} {\bibinfo {author} {\bibfnamefont {W.}~\bibnamefont
			{McGrew}}, \bibinfo {author} {\bibfnamefont {X.}~\bibnamefont {Zhang}},
		\bibinfo {author} {\bibfnamefont {R.}~\bibnamefont {Fasano}}, \bibinfo
		{author} {\bibfnamefont {S.}~\bibnamefont {Sch{\"a}ffer}}, \bibinfo {author}
		{\bibfnamefont {K.}~\bibnamefont {Beloy}}, \bibinfo {author} {\bibfnamefont
			{D.}~\bibnamefont {Nicolodi}}, \bibinfo {author} {\bibfnamefont
			{R.}~\bibnamefont {Brown}}, \bibinfo {author} {\bibfnamefont
			{N.}~\bibnamefont {Hinkley}}, \bibinfo {author} {\bibfnamefont
			{G.}~\bibnamefont {Milani}}, \bibinfo {author} {\bibfnamefont
			{M.}~\bibnamefont {Schioppo}},  \emph {et~al.},\ }\href@noop {} {\bibfield
		{journal} {\bibinfo  {journal} {Nature}\ }\textbf {\bibinfo {volume} {564}},\
		\bibinfo {pages} {87} (\bibinfo {year} {2018})}\BibitemShut {NoStop}%
	\bibitem [{\citenamefont {Olson}(2019)}]{Olson2019}%
  \BibitemOpen
  \bibfield  {author} {\bibinfo {author} {\bibfnamefont {J.}~\bibnamefont
  {Olson}},\ }
  \ \href@noop {} {Ph.D. thesis},\ \bibinfo  {school} {University
  of Colorado} (\bibinfo {year} {2019})\BibitemShut {NoStop}%
    \bibitem [{\citenamefont {Wcis{\l}o}\ \emph {et~al.}(2017)\citenamefont
		{Wcis{\l}o}, \citenamefont {Morzy{\'n}ski}, \citenamefont {Bober},
		\citenamefont {Cygan}, \citenamefont {Lisak}, \citenamefont {Ciury{\l}o},\
		and\ \citenamefont {Zawada}}]{wcislo2017experimental}%
	\BibitemOpen
	\bibfield  {author} {\bibinfo {author} {\bibfnamefont {P.}~\bibnamefont
			{Wcis{\l}o}}, \bibinfo {author} {\bibfnamefont {P.}~\bibnamefont
			{Morzy{\'n}ski}}, \bibinfo {author} {\bibfnamefont {M.}~\bibnamefont
			{Bober}}, \bibinfo {author} {\bibfnamefont {A.}~\bibnamefont {Cygan}},
		\bibinfo {author} {\bibfnamefont {D.}~\bibnamefont {Lisak}}, \bibinfo
		{author} {\bibfnamefont {R.}~\bibnamefont {Ciury{\l}o}}, \ and\ \bibinfo
		{author} {\bibfnamefont {M.}~\bibnamefont {Zawada}},\ }\href@noop {}
	{\bibfield  {journal} {\bibinfo  {journal} {Nat.~Astron.}\ }\textbf
		{\bibinfo {volume} {1}},\ \bibinfo {pages} {0009} (\bibinfo {year}
		{2017})}\BibitemShut {NoStop}%
	\bibitem [{\citenamefont {Stadnik}\ and\ \citenamefont
		{Flambaum}(2015)}]{PhysRevLett.114.161301}%
	\BibitemOpen
	\bibfield  {author} {\bibinfo {author} {\bibfnamefont {Y.~V.}\ \bibnamefont
			{Stadnik}}\ and\ \bibinfo {author} {\bibfnamefont {V.~V.}\ \bibnamefont
			{Flambaum}},\ }\href {\doibase 10.1103/PhysRevLett.114.161301} {\bibfield
		{journal} {\bibinfo  {journal} {Phys.~Rev.~Lett.}\ }\textbf {\bibinfo
			{volume} {114}},\ \bibinfo {pages} {161301} (\bibinfo {year}
		{2015})}\BibitemShut {NoStop}%
\bibitem [{\citenamefont {Berengut}\ \emph {et~al.}(2018)\citenamefont
  {Berengut}, \citenamefont {Budker}, \citenamefont {Delaunay}, \citenamefont
  {Flambaum}, \citenamefont {Frugiuele}, \citenamefont {Fuchs}, \citenamefont
  {Grojean}, \citenamefont {Harnik}, \citenamefont {Ozeri}, \citenamefont
  {Perez},\ and\ \citenamefont {Soreq}}]{Berengut2018}%
  \BibitemOpen
  \bibfield  {author} {\bibinfo {author} {\bibfnamefont {J.~C.}\ \bibnamefont
  {Berengut}}, \bibinfo {author} {\bibfnamefont {D.}~\bibnamefont {Budker}},
  \bibinfo {author} {\bibfnamefont {C.}~\bibnamefont {Delaunay}}, \bibinfo
  {author} {\bibfnamefont {V.~V.}\ \bibnamefont {Flambaum}}, \bibinfo {author}
  {\bibfnamefont {C.}~\bibnamefont {Frugiuele}}, \bibinfo {author}
  {\bibfnamefont {E.}~\bibnamefont {Fuchs}}, \bibinfo {author} {\bibfnamefont
  {C.}~\bibnamefont {Grojean}}, \bibinfo {author} {\bibfnamefont
  {R.}~\bibnamefont {Harnik}}, \bibinfo {author} {\bibfnamefont
  {R.}~\bibnamefont {Ozeri}}, \bibinfo {author} {\bibfnamefont
  {G.}~\bibnamefont {Perez}}, \ and\ \bibinfo {author} {\bibfnamefont
  {Y.}~\bibnamefont {Soreq}},\ }\href {\doibase 10.1103/PhysRevLett.120.091801}
  {\bibfield  {journal} {\bibinfo  {journal} {Phys. Rev. Lett.}\ }\textbf
  {\bibinfo {volume} {120}},\ \bibinfo {pages} {091801} (\bibinfo {year}
  {2018})}\BibitemShut {NoStop}%
\end{thebibliography}

\end{document}